\documentclass[sigconf]{acmart}

\usepackage{microtype}
\usepackage{graphicx}
\usepackage{subfigure}
\usepackage[ruled,boxed]{algorithm2e}

\usepackage{amsmath}
\usepackage{booktabs}
\usepackage{multirow}
\usepackage{balance}
\usepackage{color,soul}
\sethlcolor{cyan}
\usepackage{pifont}
\SetKwInput{Input}{Input}
\SetKwProg{kwAlignment}{Alignment}{:}{}
\SetKwProg{kwInitialize}{Initialize}{:}{}
\SetKwProg{kwAssist}{Assist}{:}{}
\SetKwProg{kwLearning}{Learning Stage}{:}{}
\SetKwProg{kwPrediction}{Prediction Stage}{:}{}
\SetKwProg{kwOrganization}{OrganizationUpdate}{:}{}
\SetKwProg{kwIntialization}{Intialization}{:}{}

\DeclareMathOperator*{\argmin}{argmin}
\def\E{\mathbb{E}} 

\def\bm{\mathbf}
\def\R{\mathbb{R}}

\def\de{\overset{\Delta}{=}}

\def\L{\mathcal{L}}

\def\Loss{L}

\AtBeginDocument{%
  \providecommand\BibTeX{{%
    \normalfont B\kern-0.5em{\scshape i\kern-0.25em b}\kern-0.8em\TeX}}}

\setcopyright{acmcopyright}
\copyrightyear{2018}
\acmYear{2018}
\acmDOI{XXXXXXX.XXXXXXX}

\acmConference[Conference acronym 'XX]{Make sure to enter the correct
  conference title from your rights confirmation emai}{June 03--05,
  2018}{Woodstock, NY}
\acmPrice{15.00}
\acmISBN{978-1-4503-XXXX-X/18/06}

\begin{document}

\title{Decentralized Multi-Target Cross-Domain Recommendation for Multi-Organization Collaborations}


\author{Enmao Diao}
\email{enmao.diao@duke.edu}
\orcid{1234-5678-9012}
\affiliation{%
  \institution{Duke University}
  \city{Durham}
  \state{NC}
  \country{USA}
}

\author{Vahid Tarokh}
\email{vahid.tarokh@duke.edu}
\affiliation{%
  \institution{Duke University}
  \city{Durham}
  \state{NC}
  \country{USA}
}

\author{Jie Ding}
\email{dingj@umn.edu}
\affiliation{%
  \institution{University of Minnesota-Twin Cities}
  \city{Minneapolis}
  \state{MN}
  \country{USA}
}

\begin{abstract}
Recommender Systems (RSs) are operated locally by different organizations in many realistic scenarios. If various organizations can fully share their data and perform computation in a centralized manner, they may significantly improve the accuracy of recommendations. However, collaborations among multiple organizations in enhancing the performance of recommendations are primarily limited due to the difficulty of sharing data and models. To address this challenge, we propose Decentralized Multi-Target Cross-Domain Recommendation (DMTCDR) with Multi-Target Assisted Learning (MTAL) and Assisted AutoEncoder (AAE). Our method can help multiple organizations collaboratively improve their recommendation performance in a decentralized manner without sharing sensitive assets. Consequently, it allows decentralized organizations to collaborate and form a community of shared interest. We conduct extensive experiments to demonstrate that the new method can significantly outperform locally trained RSs and mitigate the cold start problem.
\end{abstract}

\ccsdesc[500]{Information systems~Recommender systems}
\ccsdesc[500]{Computing methodologies~Distributed artificial intelligence}

\keywords{Recommender Systems, Assisted Learning, Decentralized Machine Learning}



\maketitle

\section{Introduction}
\begin{figure*}[tb]
\centering
 \includegraphics[width=0.8\linewidth]{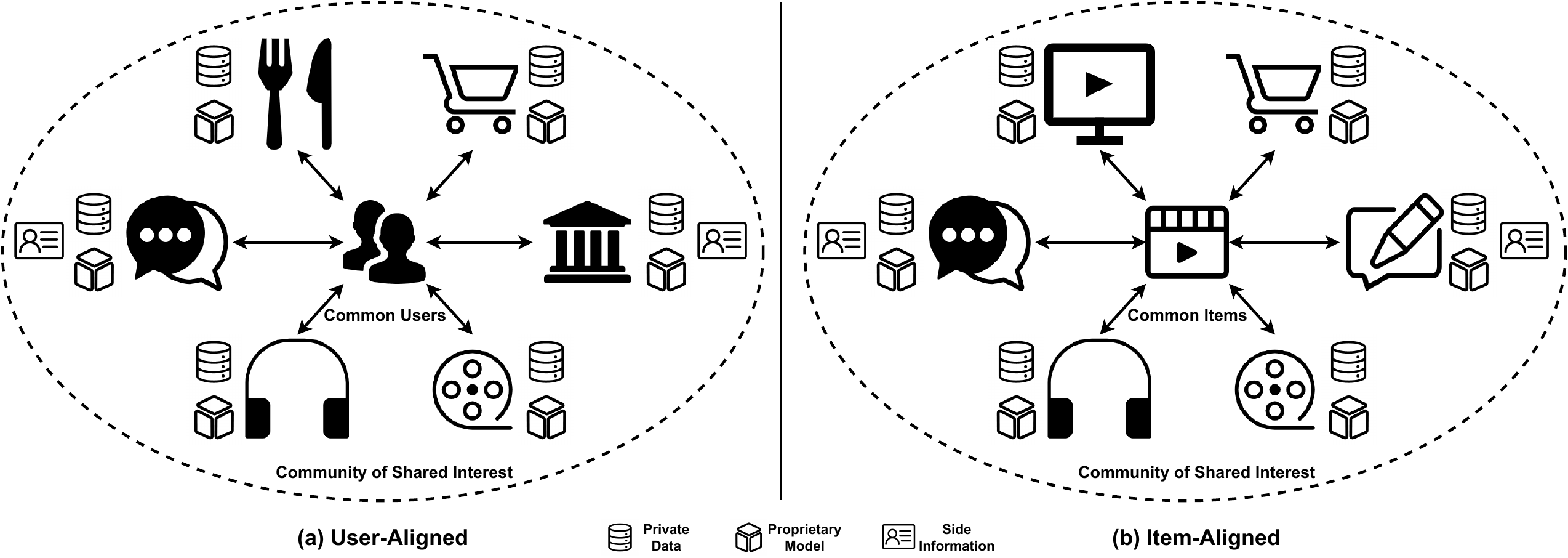}
 \caption{(a) User-Aligned (b) Item-Aligned Decentralized Multi-Target Cross-Domain Recommendation (DMTCDR) for Multi-Organization Collaborations. Decentralized organizations form a community of shared interest by leveraging the predictive power of each other without sharing their local data, model and objective functions.}
 \label{fig:mtcdr}
\end{figure*}

Recommender Systems (RSs) have become one of the most popular techniques in web applications in recent years. They can effectively extract helpful information for relevant users, e.g., recommending restaurants, videos, and e-commerce products~\cite{adomavicius2005toward, ricci2011introduction}. However, a long-standing challenge in RSs is the data sparsity problem, that the users usually interact with very few items. To address this issue, researchers have developed the direction of Cross-Domain Recommendation (CDR) to leverage ratings from a source domain where users may have relatively more information to improve the performance of a target domain~\cite{berkovsky2007cross, zhu2021cross}. A few recent works study Multi-Target Cross-Domain Recommendation (MTCDR), which aims to improve the performance of all domains simultaneously~\cite{cui2020herograph, krishnan2020transfer}. However, most existing works on this topic require that the data and models of different domains are shared, and the computation must be performed in a centralized manner~\cite{zhang2012multi,bouadjenek2018distributed,zhu2019dtcdr,cui2020herograph, krishnan2020transfer, zhu2021unified}. Since most RSs are built upon users' sensitive data, e.g., user profiles and usage history, and the model and task information are also proprietary to organizational learners~\cite{DingInfo}, collaborations among various organizations are often restricted by ethical, regulatory, and commercial constraints~\cite{zhang2019deep,zhu2021cross}. Therefore, we propose Decentralized Multi-Target Cross-Domain Recommendation (DMTCDR), which aims to simultaneously improve the performance of multiple domains in a decentralized manner. As illustrated in Figure~\ref{fig:mtcdr}, our solution provides the keystone for establishing collaborations among multiple organizations to leverage isolated data to improve recommendation performance.

Our method helps different organizations improve their recommendation performance simultaneously without sharing local data, models, and objective functions. In particular, each organization will calculate a set of `residuals' and broadcast these to other organizations. These residuals approximate the fastest direction of reducing the training loss in hindsight. Subsequently, other organizations will fit the residuals using their local data, models, and objective functions and broadcast the fitted values back to each other. Each learner will then assign weights to its peers to approximate the fastest direction of learning. The prediction will be aggregated from the fitted values. The above procedure is repeated until all organizations accomplish a sufficient level of learning. Moreover, our approach can handle explicit or implicit feedback~\cite{hu2008collaborative}, user- or item-based alignment~\cite{zhu2021cross}, and with or without side information~\cite{sun2019research}. We perform extensive experiments to demonstrate that our method significantly outperforms locally trained RSs and mitigates the cold start problem. Therefore, our method can integrate decentralized organizations to form a community of shared interest, as shown in Figure~\ref{fig:mtcdr}.
Our main contributions are as follows.

\begin{itemize}
\item We present a new recommendation framework Decentralized Multi-Target Cross-Domain Recommendation (DMTCDR), which can simultaneously improve the recommendation performance of multiple decentralized organizations without sharing their local data, models, or objective functions.

\item 
We propose a new decentralized learning algorithm named Multi-Target Assisted Learning (MTAL) with a new AutoEncoder-based RS called Assisted AutoEncoder (AAE). Our method exchanges information from various decentralized organizations by fitting pseudo-residuals with local data and models. It also covers broad application scenarios, including explicit or implicit feedback, user- or item-based alignment, and with or without side information.  

\item We conduct extensive experiments and demonstrate that our method can significantly outperform locally trained RSs and mitigates the cold start problem. As a result, our approach can promote collaborations among various organizations to form a community of shared interest.
\end{itemize}



\section{Related work}
\label{sec:related}


\paragraph{Recommender Systems} Recommender Systems (RSs) predict users' preferences on items and provide personalized recommendations for users~\cite{adomavicius2005toward,ricci2011introduction}. Recommendation approaches are mainly classified into three categories~\cite{adomavicius2005toward, jannach2010recommender}, namely collaborative filtering, content-based recommendation, and hybrid systems. Specifically, collaborative filtering learns from user-item interactions, while the content-based recommendation is primarily based on side information. Hybrid systems leverage both user-item interactions and side information. Our proposed method is a hybrid Recommender System (RS) that can leverage user-item interactions, explicit feedback (e.g., user's previous ratings) or implicit feedback (e.g., browsing history), and side information.

Users usually interact with very few items~\cite{ricci2015recommender} in most realistic scenarios. To address the data sparsity problem, cross-domain recommendation (CDR)~\cite{berkovsky2007cross} has been proposed to utilize relatively more affluent information from the source domain to improve the recommendation performance in the target domain and mitigate the cold start problem~\cite{zhu2021cross}. Recent CDR methods use the latent factors obtained from a source domain for a target domain~\cite{zhao2013active, zhang2016multi, man2017cross, gao2019cross, zhu2020deep}. A natural extension of CDR is Multi-Domain Recommendation (MDR), which aims to improve the overall recommendation performance by incorporating the domain-wise information~\cite{zhang2012multi, zhu2021unified}. Then, Dual-Target CDR (DTCDR) has been proposed to improve the recommendation performance in both the source and target domain~\cite{zhu2019dtcdr}. Inspired by MDR and DTCDR, the Multi-Target CDR (MTCDR) aims to improve the recommendation performance in all domains simultaneously~\cite{cui2020herograph, krishnan2020transfer, zhu2021unified}. The existing works require centralized training with shared data and models. However, due to various ethical and regulatory constraints, decentralized organizations may not be feasible to share their data and fully collaborate to learn a shared RS. Therefore, We propose Decentralized MTCDR (DMTCDR) to simultaneously improve the performance of multiple decentralized organizations without sharing local data, models, and objective functions.

\paragraph{Assisted Learning} Assisted Learning (AL)~\cite{xian2020assisted} is a collaborative learning framework where organizations being assisted or assisting others do not share local data and models. The recently proposed Gradient Assisted Learning (GAL)~\cite{diao2021gradient} generalizes AL from a sequential protocol to parallel aggregation across multiple organizations. However, existing work only improves the performance of one sponsor while other organizations are not benefited. We develop Multi-Target Assisted Learning (MTAL) to simultaneously improve the recommendation performance of decentralized organizations based on a novel design of AutoEncoder-based RSs. In particular, each organization trains a single local AutoEncoder to simultaneously fit residuals from multiple organizations.

\section{Problem}
\label{sec:formulation}
\subsection{Recommender Systems} Let $\mathcal{U}=\{u_{1}, \ldots, u_{m}\}$ and $\mathcal{V}=\{v_{1}, \ldots, v_{n}\}$ denote respectively the set of users and items, where $m$ is the number of users and $n$ is the number of items. We have a user-item interaction or rating matrix $\mathcal{R} \in \mathbb{R}^{m \times n}$, where $r_{i,j} \in \mathcal{R}$ denotes the rating that user $u_i$ gives to item $v_j$. 
The rating matrix $\mathcal{R}$ is sparse as each user will only interact with a few items. A Recommender System $F(\cdot)$ predicts the unseen rating with all the seen ratings. For instance, collaborative filtering is a well-known example which predicts the unseen rating $\hat{r}_{i,j}$ given a pair of user and item $(u_i, v_j)$, i.e. $\hat{r}_{i,j} = F(u_i, v_j)$. The hybrid RSs can also incorporate side information such as user profile $s_{u,i} \in \mathcal{S}_u$ and item attributes $s_{v,j} \in \mathcal{S}_v$ that are associated with the user $u_i$ and item $v_j$, so that $\hat{r}_{i,j} = F(u_i, v_j, s_{u,i}, s_{v,j})$. We train RSs by minimizing an average of loss values in the form of $\Loss(\hat{r}_{i,j}, r_{i,j})$.

\subsection{Decentralized Multi-Target Cross-Domain Recommendation} Suppose that there are $K$ observed sets of data from decentralized organizations including the user sets $\left\{\mathcal{U}^{1}, \ldots, \mathcal{U}^{K}\right\}$ and the item sets $\left\{\mathcal{V}^{1}, \ldots, \mathcal{V}^{K}\right\}$. We also have $K$ observed sets of rating matrices $\left\{\mathcal{R}^{1}, \ldots, \mathcal{R}^{K}\right\}$ associating the user and item sets, where organization $k$ has $m_k$ users and $n_k$ items. Each organization $k$ can train a local RS $\tilde{F}^k(\cdot)$. However, our proposed method leverages the common users $\mathcal{U}^{k}_c$ or the common items $\mathcal{V}^{k}_c$ that appear in other sets of users or items in order to resolve the data sparsity problem and improve the recommendation performance of Decentralized Multi-Target Cross-Domain Recommendation (DMTCDR) $F^k(\cdot)$. The set common users $\mathcal{U}^{k}_c$ and items $\mathcal{V}^{k}_c$ are those shared between a pair of organizations,
\begin{align}
    \mathcal{U}^{k}_c &= \{\mathcal{U}^{k} \cap \mathcal{U}^{1}, \ldots, \mathcal{U}^{k} \cap \mathcal{U}^{K}\} \label{eq:common_1}\\
    \mathcal{V}^{k}_c &= \{\mathcal{V}^{k} \cap \mathcal{V}^{1}, \ldots, \mathcal{V}^{k} \cap \mathcal{V}^{K}\} \label{eq:common_2}.
\end{align}
As shown in Figure~\ref{fig:mtcdr}, DMTCDR can be categorized based on their data alignment. Depending on the application scenario, a \textit{user-aligned} DMTCDR leverages the common users, while an \textit{item-aligned} one leverages the common items. 
The ultimate goal is that DMTCDR $F^k(\cdot)$ can significantly outperform locally trained RSs $\tilde{F}^k(\cdot)$ for each organization $k$
\begin{align}
    \E\{ \Loss(F^k(u^k_i, v^k_j), r^k_{i,j}) \} &\ll \E\{ \Loss(\tilde{F}^k(u^k_i, v^k_j), r^k_{i,j})\}, \label{eq:goal_1}
\end{align}
where the expectation $\E$ is over test data and $\Loss(\cdot)$ is the objective function. 

We propose Multi-Target Assisted Learning (MTAL) with Assisted AutoEncoder (AAE) to achieve the above goal for decentralized organizations without sharing local rating matrix $\mathcal{R}^{k}$, user profiles $\mathcal{S}_u^{k}$, item attributes $\mathcal{S}_v^{k}$, models $F^k(\cdot)$, or objective functions $\Loss^k(\cdot)$. 

\section{Method}
\label{sec:method}
\subsection{Multi-Target Assisted Learning} We propose Multi-Target Assisted Learning (MTAL) demonstrated in Algorithm~\ref{alg:gal}, so that \textit{each organization can operate on its own local data, model, and objective function}. We describe the learning and prediction procedures in detail. 

\begin{figure*}[htbp]
\centering
 \includegraphics[width=1\linewidth]{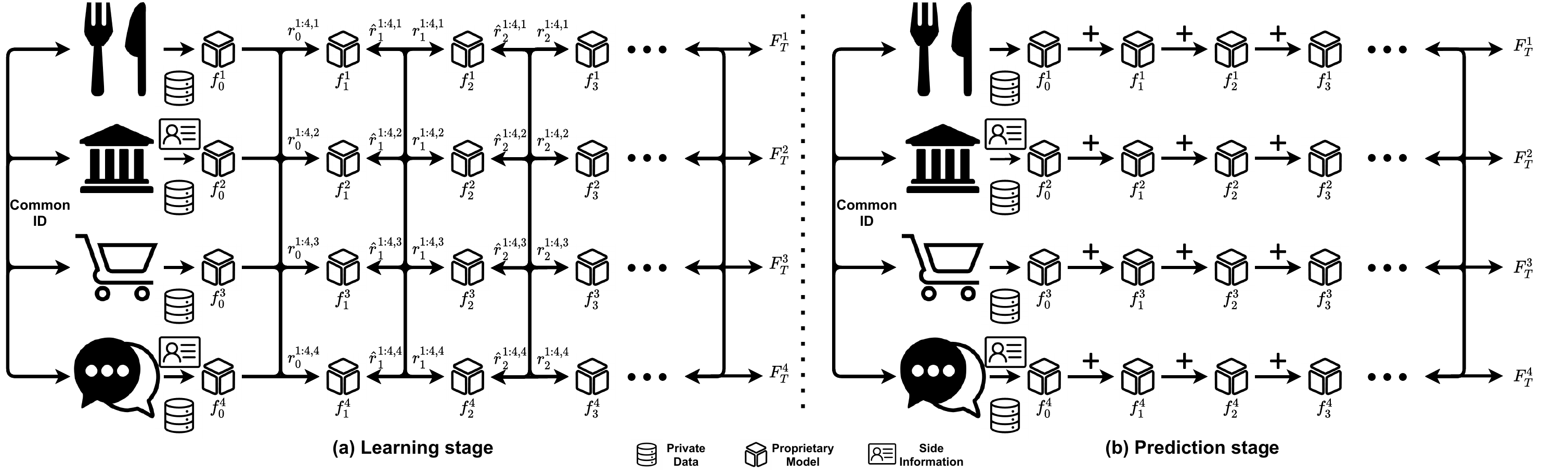}
 \caption{Learning and Prediction stages of Multi-Target Assisted Learning (MTAL). Decentralized organizations collaborate with each other and construct a set of common users or items. The organization $k$ learns local models $f_k^t$ with received pseudo-targets $r_t^{1:K, k}$ and predicted outputs $\hat{r}_t^{1:K, k}$ from all the organizations $1$ to $K$.}
 \label{fig:mtal}
\end{figure*}

\label{sec:alg}
In the beginning, each organization $k$ coordinates with other organizations to construct the set of common users $\mathcal{U}^{k}_c$ or items $\mathcal{V}^{k}_c$, depending on whether the system is user-aligned or item-aligned. During the Learning stage, each organization initializes with an unbiased base model $F^k_0(\mathcal{R}^k) = \E_n(\mathcal{R}^k) \de B^{-1} \sum \bm{r}^k_{i} \in \mathbb{R}^{n_k}$. For explicit feedback, the base model is the average ratings, where $B$ is the number of ratings of each item at the organization $k$. For the implicit feedback, the base model is the popularity estimates, where $B$ is the number of users at the organization $k$. Each organization computes its own `pseudo residuals' $r^{k,k}_t$ and broadcast its common residuals $r^{k, l}_{t}$ to another organization $l$ at every assistance round $t$, where
\begin{align*}
    r^{k,k}_t &= - \biggl[\frac{\partial \Loss_k\bigl(F_{t-1}^k(\mathcal{R}^k), \mathcal{R}^k \bigr)}{\partial F_{t-1}^k(\mathcal{R}^k)}\biggr], \\ r^{k,l}_{t} &= \biggl\{r^{k,k}_{t,i}, i \in \mathcal{U}^{k} \cap \mathcal{U}^{l} \biggr\},
\end{align*}
$\Loss^k(\cdot)$ is the overarching loss function used by organization $k$, and $F^k_{t-1}(\mathcal{R}^k)$ is the output from the previous assistance round. Here, the superscript of $r^{i,j}$ means the organization $i$ transmits the residuals to the organization $j$, or the organization $j$ receives from the organization $i$. In Figure~\ref{fig:mtal}, we let $r^{1:K,k}$ denote all the received residuals of organization $k$ from organizations from $1$ to~$K$.

\begin{algorithm*}[t]
\SetAlgoLined
\DontPrintSemicolon
\Input{$K$ decentralized organizations, organization $k$ holding rating matrix $\mathcal{R} \in \mathbb{R}^{m_k \times n_k}$, local model $f^k(\cdot)$, gradient assistance weights $w_k$, gradient assisted learning rate $\eta_k$, overarching loss function $L_k$, local loss function $\ell_k$, and the total number of assistance rounds $T$.}
\kwLearning{}{
\kwAlignment{}{
Construct the set of common users $\mathcal{U}^{k}_c$ or items $\mathcal{V}^{k}_c$
}
\kwIntialization{}{
Let $t=0$, and initialize $F^k_0(x) = \E_n(\mathcal{R}^k)$ (where $\E_n$ denotes the sample average)
}
\For{\textup{assistance round $t$ from $1$ to $T$}}{
Compute pseudo-residuals $r^{k,k}_t$ and broadcast common pseudo-residuals $r^{k,l}_{t}$ to other organizations\\
$r^{k,k}_t = - \left[\frac{\partial \Loss_k\left(F_{t-1}^k(\mathcal{R}^k), \mathcal{R}^k \right)}{\partial F_{t-1}^k(\mathcal{R}^k)}\right], \, r^{k,l}_{t} = \{r^{k,k}_{t,i}, i \in \mathcal{U}^{k} \cap \mathcal{U}^{l}\}$\\
\For{\textup{organization $m$ from $1$ to $M$} in parallel}{
Aggregates pseudo-residuals and construct pseudo-targets $\hat{\mathcal{R}}^{k}_t = \{r^{1,k}_{t}, \ldots, r^{K,k}_{t}\} \in \mathbb{R}^{m_k \times n}$\\
Fit local AAE and broadcast the common predicted outputs $\hat{r}^{k,l}_{t}$ to other organizations \\
$f^k_t = \argmin_{f^k} \ell^k \left(f^k(\mathcal{R}^k), \hat{\mathcal{R}}^k\right), \, \hat{r}^{k,k}_{t} = f^k_t(\mathcal{R}^k), \, \hat{r}^{k,l}_{t} = \{\hat{r}^{k,k}_{t,i} , i \in \mathcal{U}^{k} \cap \mathcal{U}^{l}\}$
}

Optimize the gradient assistance weights 
$w^k_t = \argmin_{w \in P_M} \ell_k \left(\sum_{j=1}^{K} w_j \hat{r}^{j,k}_{t},  \hat{\mathcal{R}}^k\right)$\\
Line search for the gradient assisted learning rate 
$\eta^k_t = \argmin_{\eta^k \in \R} \L_k\Big(F^k_{t-1}(\mathcal{R}^k) + \eta^k\sum_{j=1}^{K}w^j_t \hat{r}^{j,k}_{t}, \mathcal{R}^k\Big)$\\
$F^k_t(\mathcal{R}^k) = F^k_{t-1}(\mathcal{R}^k) + \eta^k_t\sum_{j=1}^{K}w^j_t \hat{r}^{j,k}_{t}$
}
}
\kwPrediction{}{
Gather predictions $\hat{r}^{j,k}_{t} = f^k_t(\mathcal{R}^k)$, $t=1,\ldots,T$ from each organization $j$, $j=1,\ldots,K$\\
Predict with $F^T(\mathcal{R}^k) \de F^{0}(\mathcal{R}^k) + \sum_{t=1}^T \eta_t^k\sum_{j=1}^{K}w^j_t \hat{r}^{j,k}_{t}$\\
}
Return $F^T(\mathcal{R}^k)$
\caption{MTAL: Multi-Target Assisted Learning}
\label{alg:gal}
\end{algorithm*}



Then, each organization aggregates its own residuals together with the received common residuals from other organizations into `pseudo targets' $\hat{\mathcal{R}}^{k}_t = \{r^{1,k}_{t}, \ldots, r^{K,k}_{t}\} \in \mathbb{R}^{m_k \times n}$. Note that $\hat{\mathcal{R}}^{k}_t$ is a sparse matrix of size $m_k \times n$. The dimension of items increases from $n_k$ to $n$, because the received common residuals from other organizations introduce unobserved targets of items. Next, each organization will fit a local model $f^k_t$ with the pseudo targets $\hat{\mathcal{R}}^{k}_t$ and the local loss function $\ell^k(\cdot)$. Each organization will then broadcast the common predicted outputs $\hat{r}^{k,1:K}_{t}$ to organizations from $1$ to $K$. 
\begin{align}
f^k_t &= \argmin_{f^k} \ell^k \left(f^k(\mathcal{R}^k), \hat{\mathcal{R}}^k\right),\\
\hat{r}^{k,k}_{t} &= f^k_t(\mathcal{R}^k), \, \hat{r}^{k,l}_{t} = \{\hat{r}^{k,k}_{t,i} , i \in \mathcal{U}^{k} \cap \mathcal{U}^{l}\} .
\end{align}
Subsequently, each organization can train suitable gradient assistance weights $w_k$ to aggregate received outputs and gradient assisted learning rate $\eta_k$ in minimizing the overarching loss,
\begin{align}
w^k_t &= \argmin_{w \in P_M} \ell_k \left(\sum_{j=1}^{K} w_j \hat{r}^{j,k}_{t},  \hat{\mathcal{R}}^k\right)\\
\eta^k_t &= \argmin_{\eta^k \in \R} \L_k\Big(F^k_{t-1}(\mathcal{R}^k) + \eta^k\sum_{j=1}^{K}w^j_t \hat{r}^{j,k}_{t}, \mathcal{R}^k\Big)
\end{align}
where $P_K = \{w \in \R^K: \sum_{k=1}^K w_k=1, w_k\geq 0\}$ denotes the probability simplex. Finally, the output of organization $k$ at round $t$ denoted as $F^k_t(\mathcal{R}^k)$ is the aggregation of predicted outputs from all participating organizations and the output from previous round $F^k_{t-1}(\mathcal{R}^k)$, where
\begin{align}
F^k_t(\mathcal{R}^k) = F^k_{t-1}(\mathcal{R}^k) + \eta_t^k\sum_{j=1}^{K}w^j_t \hat{r}^{j,k}_{t}.
\end{align}
The above procedure can be iterated for a total of $T$ assistance rounds until we obtain a satisfactory performance (e.g., on validation data).   
Our algorithm involves two kinds of loss functions, namely a local loss function $\ell_k(\cdot)$ for fitting the pseudo-targets $\hat{\mathcal{R}}^{k}$, and an overarching loss function $\Loss_k(\cdot)$ for fitting the ratings $\mathcal{R}^k$ in hindsight. Since fitting the pseudo-targets is a regression problem, it is standard to let $\ell_k(\cdot)$ be $\ell_1$- or $\ell_2$-norm. We choose different overarching loss functions depending on whether the ratings are explicit or implicit feedback. We use regression loss functions such as the $\ell_2$-norm for explicit feedback and classification loss functions such as binary cross-entropy for implicit feedback.

In the Prediction stage, each organization will predict outputs from their local models $f_t^k$ for all assistance rounds from $1$ to $T$. The predicted results will be broadcast to other organizations, which will aggregate them with gradient assistance weights $w_{1:T}$ and gradient assisted learning rate $\eta_{1:T}$, to eventually produce an overarching prediction $F^T(x)$ which is implicitly operated on $\mathcal{R}$. Compared with the Learning stage, the Prediction stage does not require synchronization of each assistance round because we can operate all local models across $T$ rounds before broadcasting the outputs.

It is worth noting that the Learning and Prediction stage of MTAL does not require the sharing of anyone's local data, models, or objective functions. The `pseudo-targets' allows the organizations to fit their own local data and also the residuals of other organizations. As a result, each organization can improve its local recommendation performance by leveraging the predictive power of others. The proposed MTAL algorithm develops AL in two ways. First, we generalize AL from a single-target to a multi-target learning framework. In particular, AL~\cite{xian2020assisted} and GAL~\cite{diao2021gradient} assume multiple assistors to help a single sponsor. Our MTAL method fits multiple targets of all organizations with a single local RS. Furthermore, each organization can optimize its own gradient assistance weights and learning rate to avoid negative transfer from other organizations~\cite{zhu2021cross}. As a result, MTAL can simultaneously improve the performance of all participating organizations. 



\subsection{Assisted AutoEncoder}
\begin{figure}[t]
\centering
 \includegraphics[width=1\linewidth]{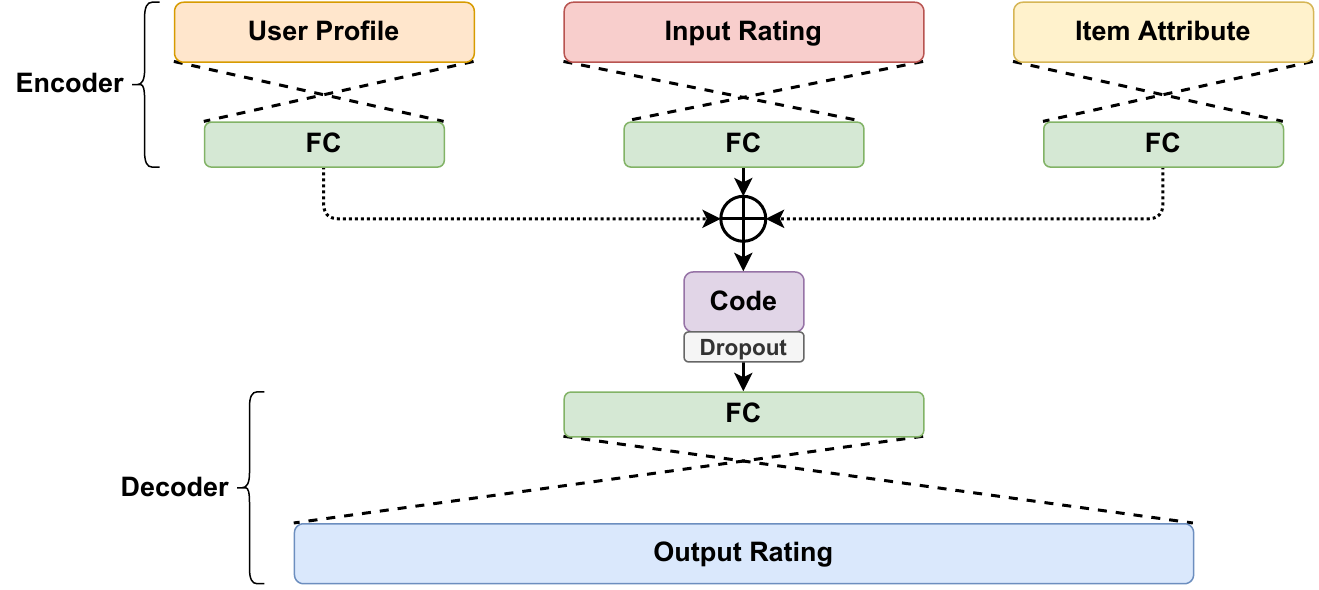}
 \vspace{-0.4cm}
 \caption{Assisted AutoEncoder (AAE) incorporates side information by summing up the codes encoded with two additional encoders corresponding to the user profile and item attribute. For user-based AAE in the organization $k$, the dimension of input ratings is the number of items of $k^{\text{th}}$ organization $n_k$, while the dimension of output ratings is the total number of items across all organizations $n$.}
 \label{fig:aae}
 \vspace{-0.2cm}
\end{figure}

A series of AutoEncoder-based RSs has found many successful applications~\cite{sedhain2015autorec,li2015deep,strub2015collaborative,liang2018variational,strub2016hybrid,ferreira2020recommendation}. An Autoencoder (AE) is a neural network consisting of an encoder $z=E(x) : R^{d_{\text{in}}} \rightarrow R^d$ and a decoder $\hat{x}=D(x) : R^d \rightarrow R^{d_{\text{out}}}$, where $d_{\text{in}}$ and $d_{\text{out}}$ are dimensions of the input vector $x$ and output vector $\hat{x}$, respectively. Unlike Collaborative Filtering (CF), the input of AutoEncoder-based RSs is not a pair of user $u_i$ and item $v_j$. Instead, the input of user-based AE is a sparse vector $\bm{r}_i = (r_{i,1}, \ldots, r_{i, n}) \in \mathbb{R}^n$ which represents the ratings of user $u_i$ giving to all items $v_1 \ldots v_n$. Similarly, the input of item-based AE is $\bm{r}_j = (r_{1,j}, \ldots, r_{m, j}) \in \mathbb{R}^m$. The encoder transforms the observed vector into a dense lower-dimensional code, namely $\bm{z}_i = E(\bm{r}_i)$ or $\bm{z}_j = E(\bm{r}_j)$. The decoder produces the output vector $\hat{\bm{r}}_i = D(E(\bm{r}_i)) \in \mathbb{R}^n$ and $\hat{\bm{r}}_j = D(E(\bm{r}_j)) \in \mathbb{R}^m$, where $n$ is the number of items. Therefore, the dimension of the input vector is the same as that of the output vector for AutoEncoder-based RSs. 




Assisted AutoEncoder (AAE) has a unique model architecture specifically designed for applying the MTAL algorithm, which is different from the ordinary AE. Suppose there are $K$ organizations, and each organization can learn its own AAE. Similar to AE, the input of a user-based AAE is a sparse vector $\bm{r}^k_i = (r^k_{i,1}, \ldots, r^k_{i, n_k}) \in \mathbb{R}^{n_k}$ that represents the ratings of user $u^k_i$ giving to all items $v^k_1 \ldots v^k_{n_k}$, where $n_k$ is the number of items in organization $k$. The key difference between AE and our proposed AAE is that the output vector of a user-based AAE has the dimension of the \textit{total} number of items of all organizations, i.e., $\hat{\bm{r}}^k_i = D(E(\bm{r}^k_i)) \in \mathbb{R}^n$. Specifically,  $n$ is the total number of items across all organizations because our proposed MTAL algorithm requires local RSs to fit the pseudo-targets of all organizations.


We illustrate our proposed AAE in Figure~\ref{fig:aae}. Both the encoder and decoder consist of Fully Connected (FC) layers. We use $\tanh(\cdot)$ as our nonlinear activation function~\cite{kuchaiev2017training,ferreira2020recommendation}. We adopt Dropout~\cite{srivastava2014dropout} at the encoded space as suggested by~\cite{kuchaiev2017training}. We also consider side information of domain $k$ such as user profile $\mathcal{S}^k_u \in \mathbb{R}^{m_k \times d^k_{s,u}}$ and item attributes $\mathcal{S}^k_v \in \mathbb{R}^{n_k \times d^k_{s,v}}$, where $d^k_{s,u}$ and $d^k_{s,v}$ denote the feature dimension of user profile and item attribute at domain $k$, respectively. The dimension of output ratings is much larger than the input ratings because we will fit the pseudo-targets of all organizations with the MTAL algorithm. Our proposed AAE can extend the scope of standard AE for DMTCDR by leveraging our proposed MTAL algorithm.

AAE is designed to generate the ratings of all organizations. Our MTAL algorithm requires that the local RSs can use the rating $r_{i,j}$ of a pair of user and item $(u_i, v_j)$ to predict the ratings $\hat{r}_{p,q}$ of other pairs of user and item, i.e.$p \neq i$ and $q \neq j$. Recall that classical CF predicts the unseen rating $\hat{r}_{i,j}$ given a single pair of user and item $(u_i, v_j)$. CF takes user-item pairs as the input, which cannot predict other user-item pairs. Thus, it is not suitable to integrate CF with our MTAL algorithm. Fortunately, an AutoEncoder-based RSs takes in all the available ratings of a user $u_i$, i.e. $(r^k_{i,1}, \ldots, r^k_{i, n_k})$ and predicts the ratings of all the unseen items, i.e. $(\hat{r}^k_{i,1}, \ldots, \hat{r}^k_{i, n_k})$. Therefore, AutoEncoder-based RSs are naturally compatible with our proposed MTAL algorithm. In particular, AE treats the rating matrix $\mathcal{R}$ as tabular data. The rating matrix's rows and columns are data samples and feature spaces. Local AutoEncoder-based RSs can be viewed as each organization holding a subset of the feature space. Specifically, user-based AE treats users as data samples and items as feature space, while item-based AE treats items as data samples and users as feature space. It is worth mentioning that the proposed MTAL algorithm is not limited to AutoEncoder-based RSs. Any models that can use the rating $r_{i,j}$ to predict the ratings $\hat{r}_{p,q}$ are also compatible with the proposed MTAL algorithm.



\vspace{-0.1cm}
\section{Experiments}
\label{sec:exp}
\subsection{Experimental Setup}
\label{sec:setup}
We conduct experiments with commonly used benchmark datasets, including MovieLens1M (ML1M), Douban, and Amazon datasets~\cite{harper2015movielens,zhu2020graphical,ni-etal-2019-justifying}. We split items of ML1M dataset according to $K=18$ movie genres, Douban dataset according to `book', `movie' and `music' domains, and Amazon dataset according to `Books', `Digital Music', `Movies and TV', and `Video Games' domains. The side information of ML1M includes users' age, gender, and occupation, and the side information of Douban is the users' living place. The summary statistics of each dataset can be found in Table~\ref{tab:data}. We compare the proposed DMTCDR with three recommendation baselines, including `Joint', MTCDR~\cite{singh2008relational}, and `Alone'. `Joint' denotes the scenario where all the data are held by one organization, and the RS is trained in a centralized manner. MTCDR is based on cross-domain Collaborative Filtering, where the embeddings of users or items are shared across multiple organizations. Compared with the proposed DMTCDR, `Joint' and MTCDR require the RS to be trained in a centralized manner. `Alone' denotes the case where organizations train local RSs and thus do not leverage the interaction among multiple organizations. 

Apart from learning baselines, we experiment various kinds of backbone recommendation models including the unbiased Base model described in Section~\ref{sec:alg}, classical Matrix Factorization (MF)~\cite{rendle2020neural}, Multi-Layer Perceptron (MLP)~\cite{he2017neural}, Neural Collaborative Filtering (NCF)~\cite{he2017neural}, and AutoEncoder (AE)~\cite{ferreira2020recommendation}. It is worth noting that each recommendation baseline requires different backbone recommendation models. In particular, we experiment with `Joint' and `Alone' baselines with all backbone models, MTCDR with backbone models based on collaborative filtering (e.g., MF, MLP, NCF), and DMTCDR with AAE. We use Root Mean Squared Error (RMSE) to evaluate explicit feedback and Normalized Discounted Cumulative Gain at rank position $10$ (NDCG@$10$) to evaluate implicit feedback. $\downarrow$ indicates the smaller the better, while $\uparrow$ indicates the larger the better. The `Improvement' is computed from the best result of the `Alone' baseline. We conduct four random experiments with different seeds. The standard errors of results are shown in the figures. Details of the experimental setup and further experimental results can be found in the Appendix. 


\begin{table}[tbp]
\centering
\caption{Summary of statistics of ML1M, Douban, and Amazon datasets. Each dataset contains $m$ users and $n$ items. $d^k_{s,u}$ represents the dimension of the side information.}
\label{tab:data}
\vspace{-0.2cm}
\resizebox{0.7\columnwidth}{!}{
\begin{tabular}{@{}cccccc@{}}
\toprule
Dataset & $m$   & $n$   & $d^k_{s,u}$ & $M$ & Sparsity \\ \midrule
ML1M    & 6040  & 3706  & 30          & 18  & 96.0\%     \\
Douban  & 2570  & 11361 & 35          & 3   & 97.5\%     \\
Amazon  & 15628 & 6946  & N/A         & 4   & 99.8\%     \\ \bottomrule
\end{tabular}
\vspace{-0.4cm}
}
\end{table}

\subsection{Experimental Results}
\begin{figure}[tbp]
\centering
 \includegraphics[width=1\linewidth]{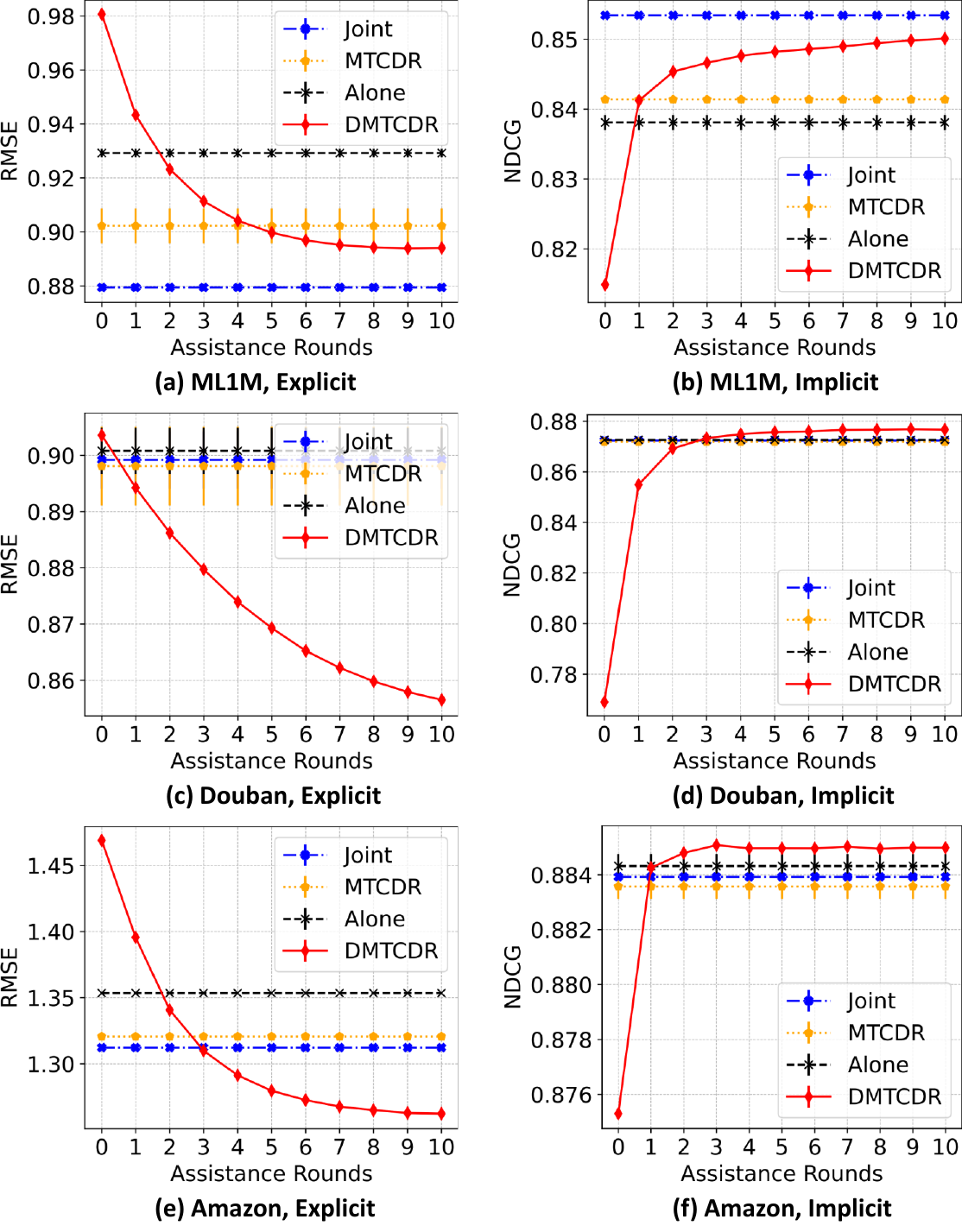}
  \vspace{-0.4cm}
 \caption{Results across all assistance rounds. User-aligned DMTCDR outperforms `Alone' baseline for both explicit and implicit feedback.}
 \label{fig:user_result}
 \vspace{-0.2cm}
\end{figure}

\paragraph{User alignment} We demonstrate the experimental results of user-aligned DMTCDR in Tables~\ref{tab:user_result}. We illustrate the evaluations of DMTCDR across all assistance rounds and the best result of each baseline in Figure~\ref{fig:user_result}. As shown in Tables~\ref{tab:user_result}, our proposed method AAE equipped with MTAL significantly outperforms all `Alone' cases for both explicit and implicit feedback with various backbone models and datasets. The results demonstrate that our decentralized framework can improve the recommendation performance of each domain simultaneously without sharing their local data, models, and objective function. Our approach also consistently outperforms the MTCDR baseline. Our method performs competitively with the `Joint' baseline for the ML1M dataset while performing better than the `Joint' baseline for Douban and Amazon datasets. However, it is worth noting that we do not expect our method can consistently outperform the `Joint' and MTCDR baselines. Because `Joint' and MTCDR baselines can be trained in a centralized manner, some more advanced methods may outperform DMTCDR~\cite{man2017cross, zhu2020deep}. Our ultimate goal is to demonstrate that it is feasible to improve the local recommendation performance by training RSs in a decentralized manner with the proposed method.

\begin{table}[tbp]
\centering
\caption{Results of ML1M, Douban, and Amazon datasets with explicit and implicit feedback. User-aligned DMTCDR improves the performance of locally trained RSs.}
\label{tab:user_result}
\vspace{-0.2cm}
\resizebox{1\columnwidth}{!}{%
\begin{tabular}{@{}cccccccc@{}}
\toprule
\multicolumn{2}{c}{Dataset} & \multicolumn{2}{c}{ML1M} & \multicolumn{2}{c}{Douban} & \multicolumn{2}{c}{Amazon} \\ \midrule
\multicolumn{2}{c}{Metric} & \multicolumn{1}{l}{RMSE$(\downarrow)$} & \multicolumn{1}{l}{NDCG$(\uparrow)$} & \multicolumn{1}{l}{RMSE$(\downarrow)$} & \multicolumn{1}{l}{NDCG$(\uparrow)$} & \multicolumn{1}{l}{RMSE$(\downarrow)$} & \multicolumn{1}{l}{NDCG$(\uparrow)$} \\ \midrule
\multirow{5}{*}{Joint} & Base & 0.981 & 0.817 & 0.904 & 0.769 & 1.469 & 0.875 \\
 & MF & 0.917 & 0.840 & 1.081 & 0.851 & 1.312 & 0.881 \\
 & MLP & 0.899 & 0.847 & 0.911 & 0.872 & 1.350 & 0.884 \\
 & NCF & 0.904 & 0.842 & 0.899 & 0.872 & 1.352 & 0.884 \\
 & AE & 0.879 & 0.853 & 0.907 & 0.869 & 1.376 & 0.883 \\ \cmidrule(l){2-8} 
\multirow{3}{*}{MTCDR} & MF & 0.925 & 0.832 & 0.955 & 0.852 & 1.321 & 0.879 \\
 & MLP & 0.908 & 0.841 & 0.905 & 0.872 & 1.341 & 0.884 \\
 & NCF & 0.902 & 0.841 & 0.898 & 0.872 & 1.339 & 0.884 \\ \midrule
\multirow{5}{*}{Alone} & Base & 0.981 & 0.815 & 0.904 & 0.769 & 1.469 & 0.875 \\
 & MF & 1.060 & 0.822 & 1.050 & 0.856 & 1.915 & 0.880 \\
 & MLP & 0.929 & 0.838 & 0.901 & 0.873 & 1.355 & 0.884 \\
 & NCF & 0.931 & 0.838 & 0.905 & 0.873 & 1.353 & 0.884 \\
 & AE & 1.158 & 0.835 & 0.905 & 0.869 & 1.528 & 0.884 \\ \cmidrule(l){2-8} 
DMTCDR & AAE & \textbf{0.894} & \textbf{0.850} & \textbf{0.857} & \textbf{0.877} & \textbf{1.262} & \textbf{0.885} \\ \midrule
\multicolumn{2}{c}{Improvement} & 3.8\% & 1.4\% & 4.9\% & 0.5\% & 6.7\% & 0.1\% \\ \bottomrule
\end{tabular}
}
\end{table}

The performance gain of DMTCDR is limited for implicit feedback. This may be due to the binary cross-entropy overarching loss used for training implicit feedback, which is less related to the metric than the mean squared error used for training explicit feedback. As illustrated in Figure~\ref{fig:user_result}, the performance gap between the `Alone' and `Joint' baseline is small for Douban and Amazon datasets because the domains of the Douban and Amazon datasets are less related than those of the ML1M dataset. In particular, local RSs can achieve satisfactory recommendation performance because they can well characterize the intra-domain interactions when the inter-domain interactions are limited. Furthermore, the results of MTCDR show that simply sharing the embeddings of items cannot effectively characterize the intra- and inter-domain interactions. Our proposed method effectively outperforms the baselines because it trains local RSs and exchanges multiple domains' predictive power by fitting their pseudo-targets.

\begin{figure}[htbp]
\centering
 \includegraphics[width=1\linewidth]{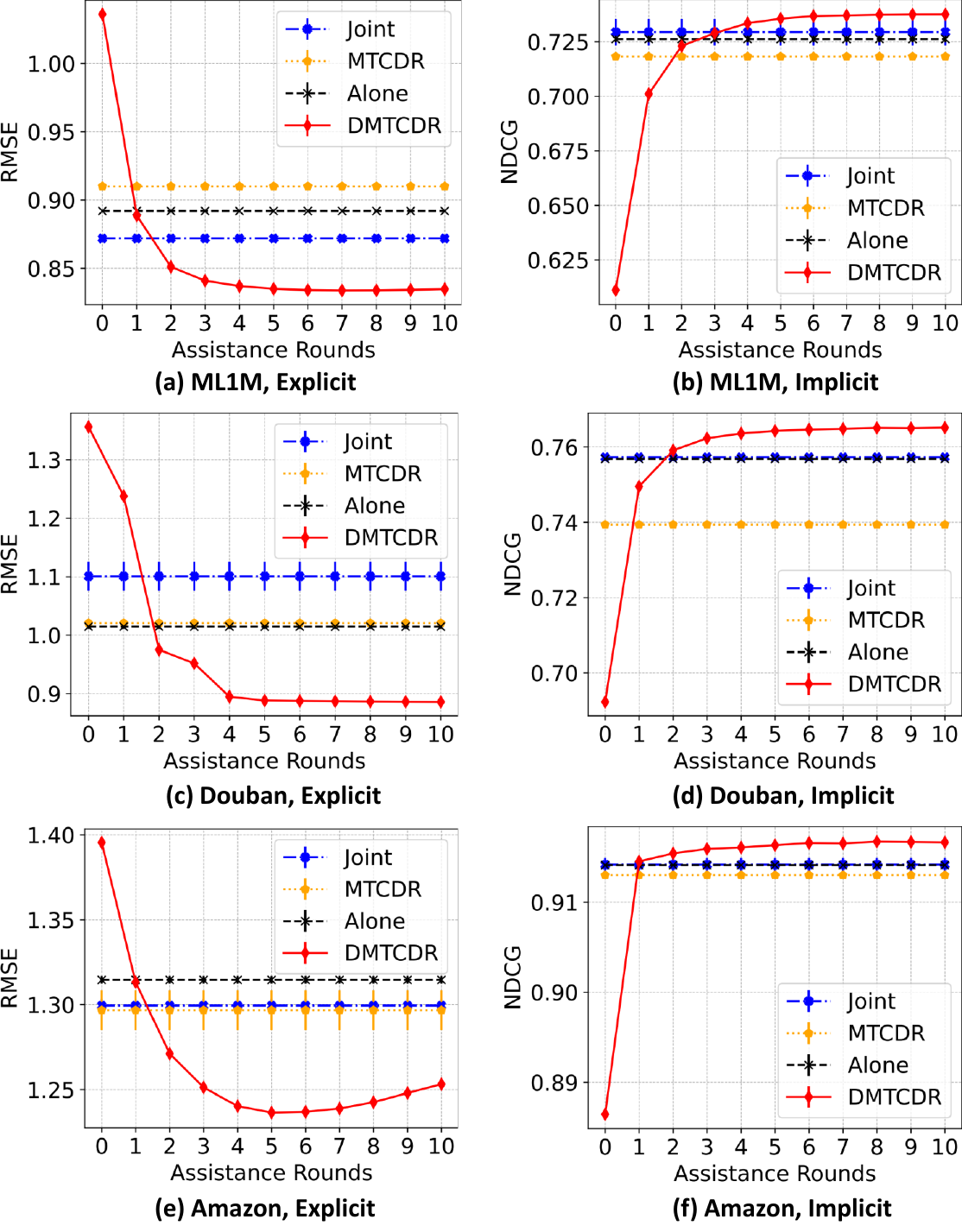}
  \vspace{-0.4cm}
 \caption{Results across all assistance rounds. Item-aligned DMTCDR outperforms `Alone' baseline for both explicit and implicit feedback.}
 \label{fig:item_result}
  \vspace{-0.2cm}
\end{figure}

\paragraph{Item alignment} We demonstrate the results of item-aligned DMTCDR in Table~\ref{tab:item_result}. We illustrate the evaluations of item-aligned DMTCDR across all assistance rounds in Figures~\ref{fig:item_result}. We randomly split users for item-aligned DMTCDR into $K=8$ domains, and each domain has roughly the same number of users. The performance gap between the `Alone' and `Joint' baseline is also reduced because the randomly constructed inter-domain interactions are much less meaningful. MTCDR fails to improve the performance of locally trained RSs as it struggles to characterize inter-domain interactions. Our proposed method performs much better than the `Alone' baseline for the item-aligned Douban dataset with explicit feedback. Consequently, DMTCDR can outperform all baselines as it can effectively characterize the intra- and inter-domain interactions.

\begin{table}[htbp]
\centering
\caption{Results of ML1M, Douban, and Amazon datasets with explicit and implicit feedback. Item-aligned DMTCDR improves the performance of locally trained RSs.}
\label{tab:item_result}
\resizebox{1\columnwidth}{!}{
\begin{tabular}{@{}cccccccc@{}}
\toprule
\multicolumn{2}{c}{Dataset} & \multicolumn{2}{c}{ML1M} & \multicolumn{2}{c}{Douban} & \multicolumn{2}{c}{Amazon} \\ \midrule
\multicolumn{2}{c}{Metric} & \multicolumn{1}{l}{RMSE$(\downarrow)$} & \multicolumn{1}{l}{NDCG$(\uparrow)$} & \multicolumn{1}{l}{RMSE$(\downarrow)$} & \multicolumn{1}{l}{NDCG$(\uparrow)$} & \multicolumn{1}{l}{RMSE$(\downarrow)$} & \multicolumn{1}{l}{NDCG$(\uparrow)$} \\ \midrule
\multirow{5}{*}{Joint} & Base & 1.036 & 0.611 & 1.356 & 0.693 & 1.396 & 0.887 \\
 & MF & 0.915 & 0.651 & 1.101 & 0.721 & 1.477 & 0.906 \\
 & MLP & 0.888 & 0.729 & 1.146 & 0.741 & 1.299 & 0.914 \\
 & NCF & 0.910 & 0.721 & 1.193 & 0.740 & 1.324 & 0.914 \\
 & AE & 0.872 & 0.728 & 1.160 & 0.757 & 1.337 & 0.913 \\ \cmidrule(l){2-8} 
\multirow{3}{*}{MTCDR} & MF & 0.915 & 0.662 & 1.021 & 0.722 & 1.459 & 0.907 \\
 & MLP & 0.910 & 0.718 & 1.027 & 0.739 & 1.297 & 0.913 \\
 & NCF & 0.910 & 0.718 & 1.134 & 0.739 & 1.342 & 0.912 \\ \midrule
\multirow{5}{*}{Alone} & Base & 1.036 & 0.611 & 1.356 & 0.693 & 1.396 & 0.886 \\
 & MF & 0.936 & 0.656 & 1.058 & 0.732 & 2.158 & 0.896 \\
 & MLP & 0.918 & 0.709 & 1.067 & 0.741 & 1.331 & 0.912 \\
 & NCF & 0.921 & 0.709 & 1.092 & 0.742 & 1.377 & 0.912 \\
 & AE & 0.892 & 0.726 & 1.015 & 0.757 & 1.315 & 0.914 \\ \cmidrule(l){2-8} 
DMTCDR & AAE & \textbf{0.834} & \textbf{0.738} & \textbf{0.886} & \textbf{0.765} & \textbf{1.236} & \textbf{0.917} \\ \midrule
\multicolumn{2}{c}{Improvement} & 6.5\% & 1.6\% & 12.7\% & 1.1\% & 6.0\% & 0.3\% \\ \bottomrule
\end{tabular}
}
\end{table}

\begin{figure}[tbp]
\centering
 \includegraphics[width=1\linewidth]{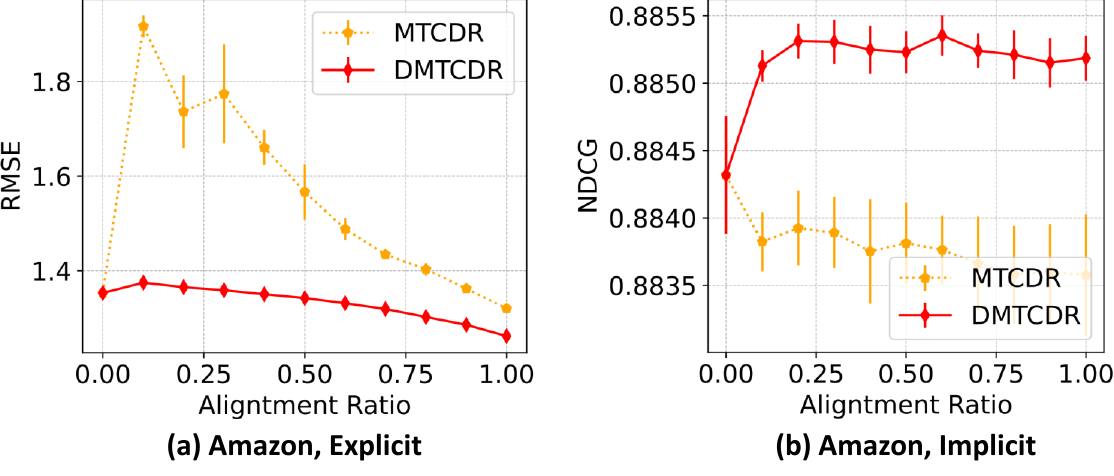}
 \caption{Results of partial alignment with Amazon dataset. DMTCDR outperforms MTCDR and `Alone' baselines when the alignment ratio increases.}
 \label{fig:partial_result_amazon}
\end{figure}

\paragraph{Partial alignment} In our previous experiments, we assume all users or items of various organizations are \textit{completely} aligned for user-aligned or item-aligned DMTCDR. As mentioned in Equations~\ref{eq:common_1} and~\ref{eq:common_2}, the common users or items are shared between a pair of domains. Specifically, two domains can share a subset of their users or items. To study the impact of alignment, we conduct an ablation study of alignment ratio. We assume that all organizations have part of users aligned according to an alignment ratio. We demonstrate the results of partial alignment in Figure~\ref{fig:partial_result_amazon}. It is worth mentioning that when the alignment ratio equals zero, the result is reduced to the best result of the `Alone' baseline. The best result of `Alone' includes all backbone models, but MTCDR and DMTCDR are limited to a subset of backbone models as described in Section~\ref{sec:setup}. As a result, MTCDR and DMTCDR may perform worse than the `Alone' baseline when the alignment ratio is small. Our results demonstrate that DMTCDR outperforms MTCDR across all alignment ratios and outperforms the `Alone' baseline when the alignment ratio is large for the Amazon dataset. It indicates that our decentralized recommendation framework is robust enough to improve the performance of partially aligned domains.

\begin{table*}[tbp]
\centering
\caption{Domain-wise results of ML1M dataset. DMTCDR outperforms the `Alone' baseline with various backbone models.}
\label{tab:ml1m_domain}
\vspace{-0.2cm}
\resizebox{2.1\columnwidth}{!}{
\begin{tabular}{@{}ccccccccccccccccccccc@{}}
\toprule
\multicolumn{3}{c}{Domain} & Action & Adventure & Animation & Children's & Comedy & Crime & Documentary & Drama & Fantasy & Film-Noir & Horror & Musical & Mystery & Romance & Sci-Fi & Thriller & War & Western \\ \midrule
\multicolumn{3}{c}{$n_k$} & 235 & 130 & 48 & 95 & 791 & 89 & 103 & 1098 & 24 & 28 & 235 & 52 & 51 & 226 & 119 & 258 & 74 & 50 \\ \midrule
\multirow{6}{*}{RMSE$(\downarrow)$} & \multirow{5}{*}{Alone} & Base & 0.973 & 0.980 & 1.002 & 1.012 & 1.001 & 0.923 & 0.984 & 0.952 & 1.026 & 0.857 & 1.095 & 0.994 & 0.958 & 0.973 & 1.005 & 0.952 & 0.944 & 0.985 \\
 &  & MF & 0.943 & 1.085 & 1.444 & 1.173 & 0.950 & 1.215 & 1.803 & 0.901 & 1.490 & 1.885 & 1.102 & 1.450 & 1.863 & 1.008 & 1.020 & 0.936 & 1.328 & 1.507 \\
 &  & MLP & 0.897 & 0.938 & 0.992 & 0.930 & 0.933 & 0.947 & 1.030 & 0.893 & 1.030 & 0.912 & 0.996 & 0.975 & 0.982 & 0.939 & 0.954 & 0.896 & 0.975 & 0.972 \\
 &  & NCF & 0.899 & 0.934 & 0.993 & 0.929 & 0.938 & 0.942 & 1.053 & 0.896 & 1.024 & 0.936 & 0.991 & 0.983 & 0.985 & 0.937 & 0.951 & 0.897 & 0.974 & 0.978 \\
 &  & AE & 1.034 & 1.153 & 1.790 & 1.305 & 1.022 & 1.265 & 1.898 & 0.985 & 2.119 & 2.360 & 1.219 & 1.667 & 1.714 & 1.058 & 1.095 & 1.013 & 1.466 & 1.797 \\ \cmidrule(l){2-21} 
 & DMTCDR & AAE & \textbf{0.871} & \textbf{0.892} & \textbf{0.920} & \textbf{0.913} & \textbf{0.911} & \textbf{0.856} & \textbf{0.933} & \textbf{0.877} & \textbf{0.933} & \textbf{0.800} & \textbf{0.989} & \textbf{0.914} & \textbf{0.876} & \textbf{0.888} & \textbf{0.915} & \textbf{0.860} & \textbf{0.860} & \textbf{0.893} \\ \midrule
\multicolumn{3}{c}{Improvement} & 2.8\% & 4.5\% & 7.3\% & 1.7\% & 2.4\% & 7.3\% & 5.2\% & 1.8\% & 8.9\% & 6.6\% & 0.1\% & 6.3\% & 8.6\% & 5.3\% & 3.8\% & 4.0\% & 8.9\% & 8.1\% \\ \midrule
\multirow{6}{*}{NDCG$(\uparrow)$} & \multirow{5}{*}{Alone} & Base & 0.705 & 0.680 & 0.651 & 0.616 & 0.761 & 0.727 & 0.754 & 0.805 & 0.546 & 0.788 & 0.590 & 0.666 & 0.662 & 0.718 & 0.678 & 0.749 & 0.736 & 0.652 \\
 &  & MF & 0.710 & 0.682 & 0.651 & 0.615 & 0.769 & 0.728 & 0.755 & 0.812 & 0.542 & 0.787 & 0.591 & 0.665 & 0.663 & 0.722 & 0.680 & 0.753 & 0.739 & 0.654 \\
 &  & MLP & 0.712 & 0.684 & 0.655 & 0.617 & 0.771 & 0.730 & 0.757 & 0.814 & 0.547 & 0.790 & 0.592 & 0.668 & 0.667 & 0.724 & 0.682 & 0.756 & 0.741 & 0.657 \\
 &  & NCF & 0.712 & 0.684 & 0.654 & 0.618 & 0.771 & 0.730 & 0.759 & 0.814 & 0.546 & 0.790 & 0.593 & 0.668 & 0.667 & 0.724 & 0.681 & 0.756 & 0.740 & 0.657 \\
 &  & AE & 0.711 & 0.679 & 0.651 & 0.610 & 0.772 & 0.728 & 0.756 & 0.815 & 0.540 & 0.790 & 0.591 & 0.664 & 0.665 & 0.722 & 0.680 & 0.757 & 0.738 & 0.656 \\ \cmidrule(l){2-21} 
 & DMTCDR & AAE & \textbf{0.714} & \textbf{0.686} & \textbf{0.656} & \textbf{0.619} & \textbf{0.777} & \textbf{0.731} & \textbf{0.759} & \textbf{0.819} & \textbf{0.548} & \textbf{0.790} & \textbf{0.593} & \textbf{0.670} & \textbf{0.668} & \textbf{0.726} & \textbf{0.684} & \textbf{0.758} & \textbf{0.742} & \textbf{0.658} \\ \midrule
\multicolumn{3}{c}{Improvement} & 0.3\% & 0.2\% & 0.2\% & 0.1\% & 0.5\% & 0.2\% & 0.0\% & 0.5\% & 0.2\% & 0.0\% & 0.1\% & 0.3\% & 0.1\% & 0.3\% & 0.3\% & 0.1\% & 0.1\% & 0.2\% \\ \bottomrule
\end{tabular}
}
\vspace{-0.2cm}
\end{table*}

\begin{figure}[tbp]
\centering
 \includegraphics[width=1\linewidth]{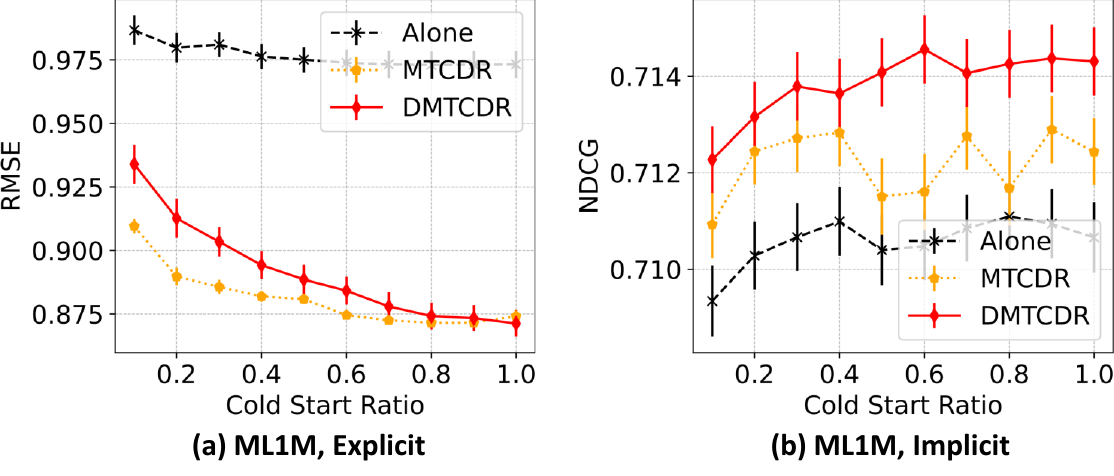}
  \vspace{-0.4cm}
 \caption{Results of cold start with ML1M dataset. `Action' is the cold start domain. DMTCDR outperforms `Alone' baseline for both explicit and implicit feedback.}
 \label{fig:cs_result}
  \vspace{-0.2cm}
\end{figure}

\paragraph{Cold start} We study the cold start problem by considering the ratio of available users of one organization, denoted as the cold start ratio. We demonstrate our results of ML1M dataset in Figure~\ref{fig:cs_result}. We choose the `Action' as the cold start organization. The cold start organization only uses the data of available users to train and align with other organizations. However, we will evaluate the performance of the cold start organization against its new users. In this case, the `Alone' baseline only works with the Base and AE backbone models because backbone models based on CF cannot train the embeddings of new users. The MTCDR baseline can also mitigate this issue because the embeddings of new users can be trained with the data from other organizations. Our proposed method can also address this issue in a decentralized manner because other organizations can assist the cold start organization through the alignment of available users. In particular, we can leverage the predictive power of other organizations by fitting their pseudo-targets. The results demonstrate that our method outperforms the `Alone' baseline and thus can mitigate the cold start problem.

\paragraph{Domain-wise improvement}
We demonstrate the domain-wise results of ML1M dataset across all user-aligned $K=18$ movie genres in Table~\ref{tab:ml1m_domain}. The results show that our proposed method DMTCDR outperforms the `Alone' baseline with various backbone models. In particular, it improves the performance of all domains simultaneously for both explicit and implicit feedback. It is worth noting that domains with a smaller number of items $n_k$ have less domain-wise improvement because the local loss functions are not reweighted according to the number of items of each domain. By reweighting the local loss functions according to the number of items of each domain, we may achieve a more fair improvement in recommendation performance over all domains.

\paragraph{Privacy enhancement}\;
Our proposed algorithm allows different domains to improve their recommendation performance without sharing local data, models, or objective functions. We consider this requirement a bottom line for protecting the privacy of DMTCDR. Nevertheless, we are aware that it is possible to apply further privacy enhancement techniques such as Differential Privacy (DP)~\cite{dwork2011differential} and Interval Privacy (IP)~\cite{DingInterval} by adding noises to the transmitted residuals~\cite{diao2021gradient}. We demonstrate the results of privacy-enhanced MTAL in Table~\ref{tab:privacy_alignment}, labeled as MTAL$_\text{DP}$ and MTAL$_\text{IP}$. The results show that privacy-enhanced MTAL can still outperform the `Alone' baseline and thus improve the local recommendation performance under privacy constraints.

\begin{table}[htbp]
\centering
\caption{Results of privacy enhancement. MTAL$_\text{DP}$ and MTAL$_\text{IP}$ represent privacy-enhanced by DP and IP.}
\label{tab:privacy_alignment}
\vspace{-0.2cm}
\resizebox{1\columnwidth}{!}{
\begin{tabular}{@{}ccccccc@{}}
\toprule
Dataset & \multicolumn{2}{c}{ML1M}              & \multicolumn{2}{c}{Douban}            & \multicolumn{2}{c}{Amazon}            \\ \midrule
Metric  & RMSE$(\downarrow)$ & NDCG$(\uparrow)$ & RMSE$(\downarrow)$ & NDCG$(\uparrow)$ & RMSE$(\downarrow)$ & NDCG$(\uparrow)$ \\ \midrule
Alone & 0.929 & 0.838 & 0.901 & 0.873 & 1.353 & 0.884 \\
MTAL$_\text{DP}$    & 0.900 & 0.850 & 0.856 & 0.875 & 1.272 & 0.885 \\
MTAL$_\text{IP}$    & 0.899 & 0.850 & 0.857 & 0.875 & 1.279 & 0.885 \\ \bottomrule
\end{tabular}
\vspace{-0.4cm}
}
\end{table}

\vspace{-0.3cm}
\section{Conclusion}
\label{sec:conclusion}
In this work, we present a new recommendation framework Decentralized Multi-Target Cross-Domain Recommendation (DMTCDR), which can simultaneously improve the recommendation performance of multiple decentralized organizations without sharing sensitive assets. Our proposed solution consists of a new decentralized learning algorithm named Multi-Target Assisted Learning (MTAL) and a new AutoEncoder-based RS called Assisted AutoEncoder (AAE). Our method covers broad application scenarios, including user- or item-based alignment, explicit or implicit feedback, and with or without side information. We conduct extensive experiments and demonstrate that our method can significantly outperform locally trained Recommender Systems (RSs) and mitigate the cold start problem. Consequently, our approach can effectively promote collaborations among various organizations to form a community of shared interest.

\section*{Acknowledgments}

This work of Enmao Diao and Vahid Tarokh was supported in part by the Office of Naval Research under grant number N00014-18-1-2244. Jie Ding was supported by the Office of Naval Research under grant number N00014-21-1-2590.



% ...




\balance

\bibliographystyle{unsrt}
\bibliography{References}

\newpage
\section*{Appendix}

\appendix

\section{Limitations and Future Work}
Our proposed work extends and applies Assisted Learning (AL) for Decentralized Multi-Target Cross-Domain Recommendation (DMTCDR) with Multi-Target Assisted Learning (MTAL) and Assisted AutoEncoder (AAE). The results show that our method can outperform baselines in most scenarios. However, obtaining a dominant advantage over all existing works can be challenging. First, we can further improve the performance with hyperparameter searching methods in practice~\cite{agtabular}. Second, we only discover AAE as one candidate backbone model for MTAL. The performance may be improved with more advanced backbone models such as graph-based RSs~\cite{wang2021graph} as long as they are compatible with MTAL.
Furthermore, the domain-wise results show that the improvement over locally trained RS is not fair enough across various domains. Finally, we do not focus on preserving privacy because AL still lacks a comprehensive analysis on this subject. Instead, we focus on decentralized computation without sharing local data, models, or objective functions. Furthermore, we also conduct experiments with privacy-enhancing techniques. Nonetheless, it is desirable to study the privacy aspect of DMTCDR further.

\section{Experimental Setup}
We train on $90\%$ of the available data for all datasets and test on the remaining. We set negative implicit feedback when the ratings are below $3.5$. Since the original Douban and Amazon datasets contain extremely sparse entries, we use users and items with more than $20$ associated ratings. We demonstrate the model architecture of backbone models used in our experiments in Table~\ref{tab:arch}. Details of hyper-parameters are included in Tables~\ref{tab:hyper}. 


\begin{table}[htbp]
\centering
\caption{The model architecture of models used in our experiments. The size embeddings of MF and NCF is $128$. The Fully Connected (FC) neural networks used in MLP and NCF have four layers of size $[128, 64, 32, 16]$. Our proposed AAE has a two-layer encoder of size $[n_k \text{ or } m_k, 256, 128]$ and a two-layer decoder of size $[128, 256, n \text{ or } m]$.}
\label{tab:arch}
\vspace{-0.2cm}
\resizebox{0.7\columnwidth}{!}{
\begin{tabular}{@{}cc@{}}
\toprule
\multicolumn{1}{l}{Model} & Architecture                   \\ \midrule
MF                        & $128$                            \\
MLP                       & $[128, 64, 32, 16]$           \\
NCF                       & $[128, 64, 32, 16]$                      \\
AAE                       & $[n_k \text{ or } m_k, 256, 128]$, $[128, 256, n \text{ or } m]$ \\ \bottomrule
\end{tabular}
\vspace{-0.4cm}
}
\end{table}

\section{Experimental Results}
\subsection{Ablation studies}
We conduct an ablation study of the gradient assisted learning rate $\eta_k$ of three datasets. As demonstrated in Table~\ref{tab:ablation}, `$\eta_k$' represents a constant gradient assisted learning rate for all domains, and the optimization of gradient assistance weights $w_k$ is disabled. In particular, the weighted average becomes an unweighted average of outputs. We observe that the optimization of $\eta_k$ for explicit feedback may result in overfitting due to a very large $\eta_k$. A moderate $\eta_k$ produce better results than a large $\eta_k$. In practice, a more reliable solution to determine $\eta_k$ is to use a validation set. We use the best choice of the gradient assisted learning rate $\eta_k$ to perform the ablation study of gradient assistance weights $w_k$. As demonstrated in Table~\ref{tab:ablation}, the impact of optimizing $w_k$ is not significant in terms of recommendation performance. It is worth noting that previous work shows that $w_k$ is beneficial against adversarial training~\cite{diao2021gradient}.

\begin{table}[htbp]
\centering
\caption{Hyperparameters used for training local models, gradient assistance weights $w_m$, and gradient assisted learning rates $\eta_m$.}
\label{tab:hyper}
\vspace{-0.2cm}
\resizebox{0.7\columnwidth}{!}{
\begin{tabular}{@{}ccccc@{}}
\toprule
\multicolumn{2}{c}{Dataset} & ML1M & Douban & Amazon \\ \midrule
\multirow{5}{*}{Local} & Batch size & 500 & 100 & 500 \\ \cmidrule(l){2-5} 
 & Epoch & \multicolumn{3}{c}{20} \\ \cmidrule(l){2-5} 
 & Optimizer & \multicolumn{3}{c}{Adam} \\ \cmidrule(l){2-5} 
 & Learning rate & \multicolumn{3}{c}{1.0E-03} \\ \cmidrule(l){2-5} 
 & Weight decay & \multicolumn{3}{c}{5.0E-04} \\ \midrule
\multirow{4}{*}{\begin{tabular}[c]{@{}c@{}}Optimize\\ $\eta_k,w_k$\end{tabular}} & Epoch & \multicolumn{3}{c}{20} \\ \cmidrule(l){2-5} 
 & Batch size & \multicolumn{3}{c}{Full} \\ \cmidrule(l){2-5} 
 & Optimizer & \multicolumn{3}{c}{L-BFGS} \\ \cmidrule(l){2-5} 
 & Learning rate & \multicolumn{3}{c}{1} \\ \midrule
\multicolumn{2}{c}{Assistance rounds} & \multicolumn{3}{c}{10} \\ \bottomrule
\end{tabular}
\vspace{-0.4cm}
}
\end{table}

\begin{table}[htbp]
\centering
\caption{Ablation study of gradient assisted learning rate and gradient assistance weights.}
\label{tab:ablation}
\vspace{-0.2cm}
\resizebox{1\columnwidth}{!}{
\begin{tabular}{@{}ccccccc@{}}
\toprule
Dataset           & \multicolumn{2}{c}{ML1M} & \multicolumn{2}{c}{Douban} & \multicolumn{2}{c}{Amazon} \\ \midrule
Metric &
  \multicolumn{1}{l}{RMSE$(\downarrow)$} &
  \multicolumn{1}{l}{NDCG$(\uparrow)$} &
  \multicolumn{1}{l}{RMSE$(\downarrow)$} &
  \multicolumn{1}{l}{NDCG$(\uparrow)$} &
  \multicolumn{1}{l}{RMSE$(\downarrow)$} &
  \multicolumn{1}{l}{NDCG$(\uparrow)$} \\ \midrule
$\eta_k=0.1$      & 0.911       & 0.841      & 0.857        & 0.853       & 1.389        & 0.885       \\
$\eta_k=0.3$      & 0.894       & 0.847      & 0.854        & 0.873       & 1.310        & 0.885       \\
$\eta_k=1.0$      & 0.889       & 0.850      & 0.852        & 0.877       & 1.262        & 0.885       \\
Optimize $\eta_k$ & 0.896       & 0.829      & 0.857        & 0.853       & 1.304        & 0.882       \\
Optimize $w_k$    & 0.894       & 0.850      & 0.856        & 0.877       & 1.333        & 0.885       \\ \bottomrule
\end{tabular}
\vspace{-0.4cm}
}
\end{table}

\subsection{Partial alignment}
\begin{figure}[htbp]
\centering
 \includegraphics[width=1\linewidth]{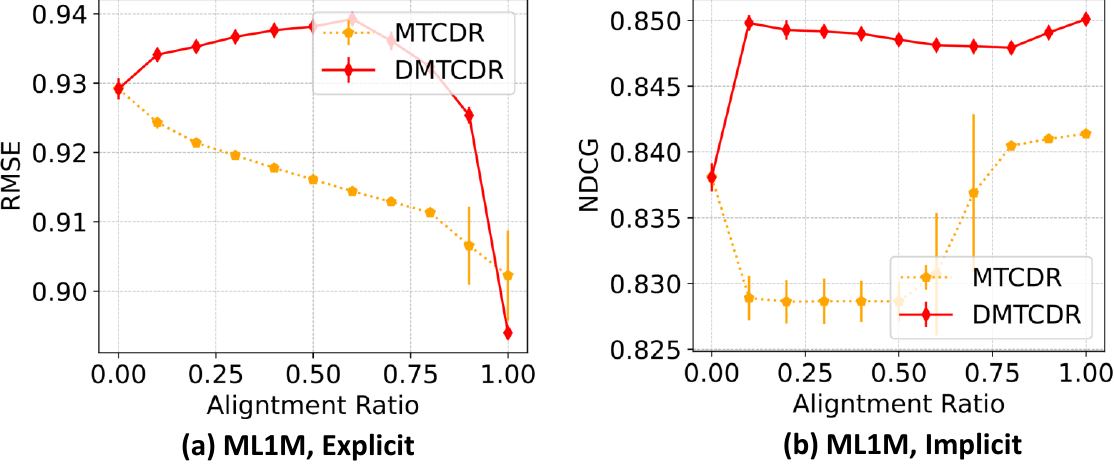}
  \vspace{-0.4cm}
 \caption{Results of partial alignment with ML1M dataset. DMTCDR outperforms MTCDR and `Alone' baselines when the alignment ratio increases.}
 \label{fig:partial_result_ml1m}
  \vspace{-0.2cm}
\end{figure}

\begin{figure}[htbp]
\centering
 \includegraphics[width=1\linewidth]{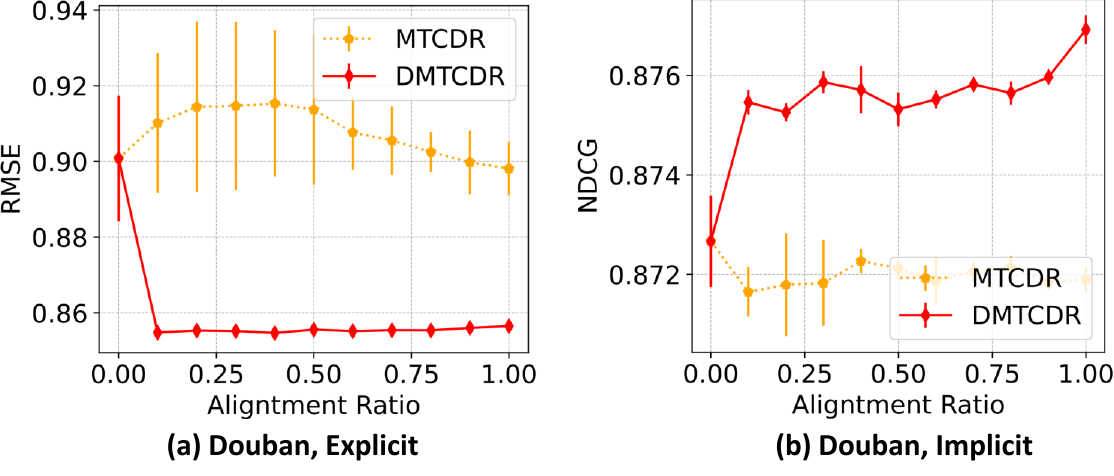}
  \vspace{-0.4cm}
 \caption{Results of partial alignment with Douban dataset. DMTCDR outperforms MTCDR and `Alone' baselines when the alignment ratio increases.}
 \label{fig:partial_result_douban}
  \vspace{-0.2cm}
\end{figure}

In Figure~\ref{fig:partial_result_ml1m} and~\ref{fig:partial_result_douban}, we demonstrate the results of partial alignment for ML1M and Douban datasets. The results demonstrate that DMTCDR performs worse than MTCDR for the ML1M dataset with explicit feedback when the alignment ratio is small. It may be because the chosen gradient assisted learning rate is not optimal for the small alignment ratio. Our method outperforms MTCDR in other experiments. The performance improvement of alignment converges at a small alignment ratio for the ML1M dataset with implicit feedback and Douban dataset with explicit feedback. Meanwhile, the result of the Douban dataset with implicit feedback continues to improve when the alignment ratio increases. Consequently, the results demonstrate that DMTCDR performs better at a large alignment ratio while potentially converging at a small one.

\subsection{Cold start}
\begin{figure}[htbp]
\centering
 \includegraphics[width=1\linewidth]{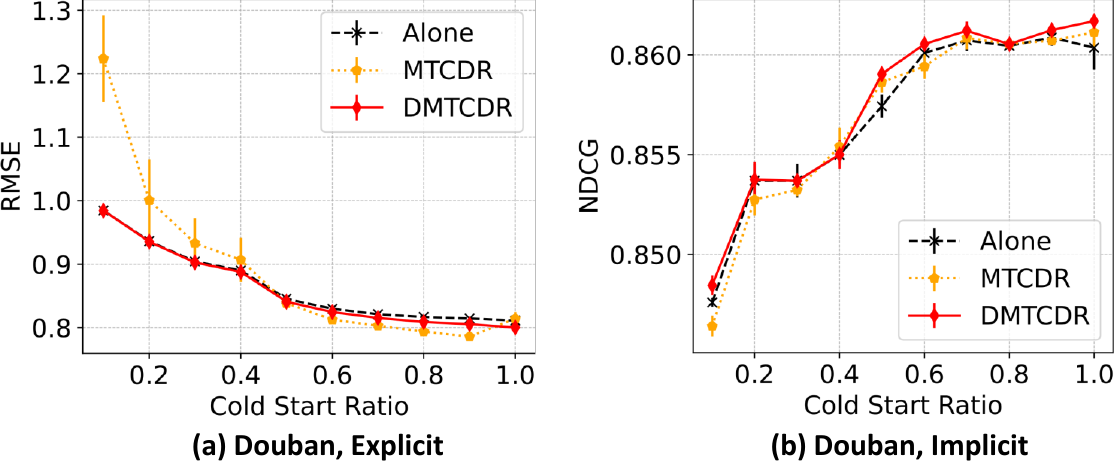}
  \vspace{-0.4cm}
 \caption{Results of cold start with Douban dataset. `Book' is the cold start domain. DMTCDR outperforms MTCDR and `Alone' baselines when the cold start ratio increases.}
 \label{fig:cs_result_douban}
  \vspace{-0.2cm}
\end{figure}

\begin{figure}[htbp]
\centering
 \includegraphics[width=1\linewidth]{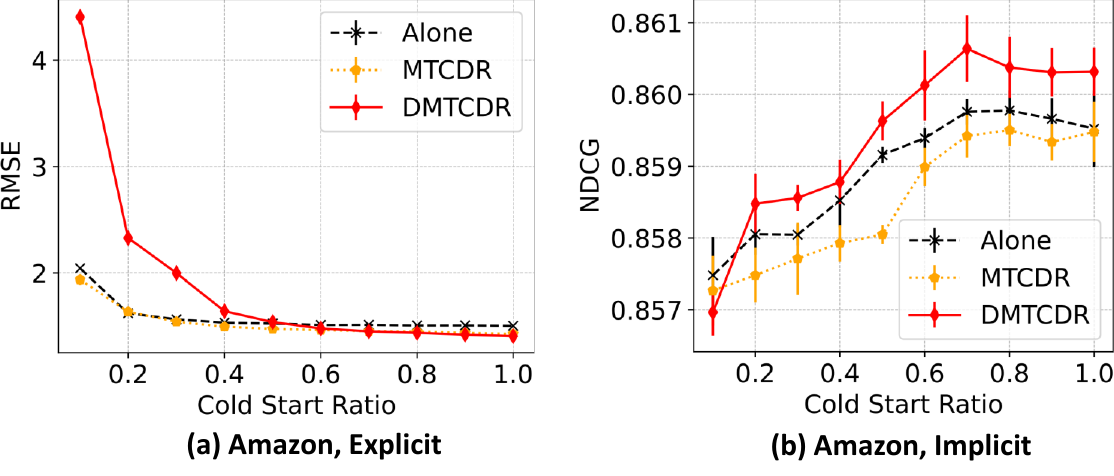}
  \vspace{-0.4cm}
 \caption{Results of cold start with Amazon dataset. `Books' is the cold start domain. DMTCDR outperforms MTCDR and `Alone' baselines when the cold start ratio increases.}
 \label{fig:cs_result_amazon}
  \vspace{-0.2cm}
\end{figure}

In Figure~\ref{fig:cs_result_douban} and~\ref{fig:cs_result_amazon}, we demonstrate the results of cold start for Douban and Amazon datasets. We choose the `Book' as the cold start organization for the Douban dataset and the `Books' as the cold start organization for the Amazon dataset, respectively. The cold start organization only uses the data of available users to train and align with other organizations. DMTCDR performs worse than MTCDR for the Amazon dataset with explicit feedback when the cold start ratio is small. It may be because the performance of the Base model, used as the initial state of DMTCDR, is poor when the cold start ratio is small. However, the performance of DMTCDR continues to improve when the cold start ratio increases. Consequently, the results demonstrate that DMTCDR performs better at a large cold start ratio while potentially performing worse than MTCDR at a small cold start ratio.

\subsection{Domain-wise improvement}
In Tables~\ref{tab:douban_domain} and~\ref{tab:amazon_domain}, we demonstrate the results of each domain of user-aligned datasets. The results show that DMTCDR improves the performance of all domains simultaneously. The results of DMTCDR for the Amazon dataset (`Digital Music') with explicit feedback perform worse than its corresponding `Alone' baseline. It may be because the performance of the AE is poor as the number of items of `Digital Music' is much smaller than that of other domains. Nonetheless, DMTCDR outperforms `Alone' baselines in other scenarios.

\begin{table}[htbp]
\centering
\caption{Domain-wise results of Douban dataset.}
\label{tab:douban_domain}
\vspace{-0.2cm}
\resizebox{0.7\columnwidth}{!}{
\begin{tabular}{@{}cccccc@{}}
\toprule
\multicolumn{3}{c}{Domain} & Book & Movie & Music \\ \midrule
\multicolumn{3}{c}{$n_k$} & 1134 & 9500 & 727 \\ \midrule
\multirow{6}{*}{RMSE$(\downarrow)$} & \multirow{5}{*}{Alone} & Base & 0.811 & 0.908 & 0.855 \\
 &  & MF & 0.847 & 1.059 & 0.963 \\
 &  & MLP & 0.817 & 0.905 & 0.860 \\
 &  & NCF & 0.847 & 0.907 & 0.879 \\
 &  & AE & 0.857 & 0.907 & 0.910 \\ \cmidrule(l){2-6} 
 & DMTCDR & AAE & \textbf{0.800} & \textbf{0.859} & \textbf{0.843} \\ \midrule
\multicolumn{3}{c}{Improvement} & 1.3\% & 5.1\% & 1.4\% \\ \midrule
\multirow{6}{*}{NDCG$(\uparrow)$} & \multirow{5}{*}{Alone} & Base & 0.846 & 0.770 & 0.843 \\
 &  & MF & 0.858 & 0.850 & 0.853 \\
 &  & MLP & 0.861 & 0.868 & 0.858 \\
 &  & NCF & 0.861 & 0.868 & 0.858 \\
 &  & AE & 0.860 & 0.865 & 0.858 \\ \cmidrule(l){2-6} 
 & DMTCDR & AAE & \textbf{0.862} & \textbf{0.873} & \textbf{0.859} \\ \midrule
\multicolumn{3}{c}{Improvement} & 0.0\% & 0.6\% & 0.1\% \\ \bottomrule
\end{tabular}
}
\end{table}

\begin{table}[htbp]
\centering
\caption{Domain-wise results of Amazon dataset.}
\label{tab:amazon_domain}
\vspace{-0.2cm}
\resizebox{1\columnwidth}{!}{
\begin{tabular}{@{}ccccccc@{}}
\toprule
\multicolumn{3}{c}{Domain} & Books & Digital Music & Movies and TV & Video Games \\ \midrule
\multicolumn{3}{c}{$n_k$} & 2241 & 343 & 3178 & 1184 \\ \midrule
\multirow{6}{*}{RMSE$(\downarrow)$} & \multirow{5}{*}{Alone} & Base & 1.632 & 1.388 & 1.339 & 1.483 \\
 &  & MF & 2.125 & 3.802 & 1.348 & 2.225 \\
 &  & MLP & 1.487 & \textbf{1.173} & 1.254 & 1.362 \\
 &  & NCF & 1.488 & 1.178 & 1.250 & 1.360 \\
 &  & AE & 1.504 & 2.767 & 1.359 & 1.588 \\ \cmidrule(l){2-7} 
 & DMTCDR & AAE & \textbf{1.406} & 1.237 & \textbf{1.144} & \textbf{1.247} \\ \midrule
\multicolumn{3}{c}{Improvement} & 5.4\% & -5.5\% & 8.5\% & 8.3\% \\ \midrule
\multirow{6}{*}{NDCG$(\uparrow)$} & \multirow{5}{*}{Alone} & Base & 0.856 & 0.906 & 0.859 & 0.832 \\
 &  & MF & 0.857 & 0.906 & 0.862 & 0.832 \\
 &  & MLP & 0.859 & 0.906 & 0.866 & 0.833 \\
 &  & NCF & 0.859 & 0.906 & 0.866 & 0.833 \\
 &  & AE & 0.860 & 0.906 & 0.866 & 0.832 \\ \cmidrule(l){2-7} 
 & DMTCDR & AAE & \textbf{0.860} & \textbf{0.906} & \textbf{0.867} & \textbf{0.833} \\ \midrule
\multicolumn{3}{c}{Improvement} & 0.1\% & 0.0\% & 0.1\% & 0.0\% \\ \bottomrule
\end{tabular}
}
\end{table}

\end{document}